\documentclass[preprint,showpacs,preprintnumbers,amsmath,amssymb]{revtex4}

\usepackage{dcolumn}
\usepackage{bm}

\begin{document}

\bibliographystyle{prsty}


\title{Topological effects, index theorem and supersymmetry \\in graphene}
\author{K.-S. Park}
\email{kpark@postech.ac.kr} \affiliation{Department of Electrical
and Computer Engineering, Pohang University of Science and
Technology, San 31 Hyoja-Dong Nam-Gu, Pohang, Kyungbuk 790-784,
Korea }

\date{\today}

\begin{abstract}

We present the electronic properties of massless Dirac fermions
characterized by geometry and topology on a graphene sheet in this
chapter. Topological effects can be elegantly illuminated by the
Atiyah-Singer index theorem. It leads to a topological invariant
under deformations on the Dirac operator and plays an essential role
in formulating supersymmetric quantum mechanics over twisted
Dolbeault complex caused by the topological deformation of the
lattice in a graphene system. Making use of the G index theorem and
a high degree of symmetry, we study deformed energy eigenvalues in
graphene. The Dirac fermion results in SU(4) symmetry as a high
degree of symmetry in the noninteracting Hamiltonian of the
monolayer graphene. Under the topological deformation the
zero-energy states emerge naturally without the Zeeman splitting at
the Fermi points in the graphene sheet. In the case of nonzero
energy, the up-spin and down-spin states have the exact high
symmetries of spin, forming the pseudospin singlet pairing. We
describe the peculiar and unconventional quantum Hall effects of the
$ n = 0 $ Landau level in monolayer graphene on the basis of the G
index theorem and the high degree of symmetry.

\end{abstract}

\pacs{73.43.-f, 75.80.=g, 71.27.+a,11.15.-q}
\keywords{Molecular Transistors, Nano-based Transistors, Graphene,
Massless Dirac fermions, Quantum Hall effect, Berry phase, Supersymmetry}

\maketitle
\section{\label{sec:level1}Introduction}

Carbon provides a fundamental material for all life and physical
science. Carbon-based systems reveal a variety of structures with a
great deal of physical properties. These physical properties result
from the dimensionality of the structures among systems with carbon
atoms. For a long time, in material science, both experimentalists
and theorists have sought for the existence of a true
two-dimensional (2D) material with the thickness of a single atom,
or a membrane of atomic thickness. This 2D material was
theoretically first studied on a monolayer of graphite by Wallace
\cite{wallace}. Experimentally in the year 2004, a group led by A.
K. Geim at the University of Manchester, U. K., realized such a 2D
material under the name of graphene \cite{castroneto,novoselov1}.

Graphene is composed of carbon atoms placed at the vertices of a two
dimensional honeycomb lattice. It is regarded as a large molecule of
carbon atoms which become strongly bound together on the sites of
the honeycomb lattice. For each carbon atom on the lattice, three of
the four outer electrons get strongly bond with its neighboring
atoms by $ \sigma $ orbitals. The 2$ p_{z} $ orbital of the fourth
electron produces a $ \pi $ bond with a neighboring carbon atom. The
$ \sigma $ bonds form the covalent structure with a honeycomb
geometry. The bond strength furnishes the flexibility and robustness
for the lattice geometry. On the other hand, the $ \pi $ bonds
generate the intrinsic electronic structure of graphene. Each $ \pi
$ bond yields the half-filled electrons of $ p $ orbital to tunnel
from a carbon atom to the neighboring one. Thus graphene should be
regarded as a many body system on which electrons can get correlated
from site to site, resulting in a rich collective behavior. The
correlated behavior can be represented by quantum effects which can
influence on graphene's electronic properties \cite{castroneto}.

The electronic structure can be described by 2D massless
relativistic fermions\cite{semenoff,haldane,novoselov2,zhang} in
graphene. The massless fermions enable us to study topological
effects on electronic properties of graphene. Topological effects
are represented by the global properties of geometrical objects
rather than their local ones. By varying the geometry, we can
produce topologically different configurations such as a sphere or a
torus on which the effective Dirac operators of massless fermions
are well defined. Described by the Dirac fermions, graphene can have
extraordinary properties of stability obtained in terms of geometry
and topology of the underlined lattice. Apart from the robust
structure of geometry, the topological properties can emerge due to
long range quantum coherence in graphene. It follows that we can
move electrons coherently through the whole graphene molecule,
resulting in its detection of geometry or topology. This allows us
to study a great deal of physical properties revealed by interplay
between geometry and topology, and quantum
effects\cite{pachos,park}.

As quantum effects, there exist unconventional quantum Hall effects
(QHE) which form a series of filling factors $ \nu = \pm 2, \pm 6,
\pm 10,\cdots $ as the four-fold degeneracy combined by spin and
sublattice valley ones \cite{novoselov2,zhang}. The energy
dispersion shows a linear spectrum by the massless Dirac fermions
with a Fermi velocity {$ v_{\mathrm{F}} \approx 10^{6} m/s $}. The
4-fold degeneracy of the Landau level (LL) is lifted into 4
sublevels in the presence of an external high magnetic field. In the
case of the tilted high magnetic field to the graphene plane, the
spin degeneracy can be lifted at the firsr LL, resulting in the
filling factor $\nu = 4$ QHE of monolayer graphene \cite{kim}.
Furthermore for bilayer graphene, the LL spectrum is composed of
eightfold degenerate states at the zero energy and fourfold ones at
finite energies under the high magnetic field. This can allows us to
observe the quantum Hall plateaus at a seris of $ \nu = \pm 4, \pm
8, \pm 12, \cdots $ \cite{geim,feldman,zhao}. The charge carriers
are chiral massive fermions which produce a parabolic energy band.
The chiral fermions offer the unconventional integer QHE of the
zero-LL anomaly which exhibits metallic behavior under the condition
of low carrier densities and high magnetic fields in contrast to the
conventional insulating phenomena
\cite{novoselov2,zhang,park,kim,geim,feldman,zhao,mccann}.

Topological configurations can produce a crucial effect on the
quantum states of a system. In particular, they can provide the
possible quantum ground states which a system can have. This
remarkable result is described in terms of the index theorem
initiated by Atiyah and Singer \cite{atiyah}. It gives the
relationship between the analytic properties of the operator and the
topological characteristic of the manifold upon which the operator
is defined. The Dirac operator can be related to topological effects
which is elegantly illuminated by the Atiyah-Singer index theorem in
graphene \cite{pachos}. It leads to a topological invariant under
deformations on a Dirac operator and plays an essential role in
formulating supersymmetric quantum mechanics (SUSY QM) on the
graphene sheet \cite{park,witten}.

In a theoretical sense, there has been at least the quantum
mechanics (QM) of particles described by both fermionic and bosonic
degrees of freedom. The SUSY QM may be hidden in the quantum
mechanics of a particle acting on a group manifold which can be
represented by a high degree of symmetry \cite{stone}. As an
example, a spin precessing in a magnetic field can have the hidden
SUSY. In particular, under the uniform magnetic field the LL for an
electron can be expressed by the spectrum of the SUSY oscillator
which is composed of fermionic and bosonic ones. It is remarkable
that this SUSY QM can possibly emerge in graphene with low carrier
concentration and high mobility. The supersymmetry is built up over
the Dolbeault complex due to the topological deformation on the
lattice in a graphene system \cite{stone,park,ezawa}.

We exploit the G-index theorem and a high degree of symmetry to
understand unusual quantum Hall effects of the $ n = 0 $ Landau
level in graphene. The Dirac fermion results in SU(4) symmetry as a
high degree of symmetry in the noninteracting Hamiltonian of the
monolayer graphene. The high symmetries in graphene sheets can not
couple to an external magnetic field. In the absence of the magnetic
field the index theorem can provide a relation between the
zero-energy state of the graphene sheet and the topological
deformation of the compact lattice. Under the topological
deformation the zero-energy states emerge naturally without the
Zeeman splitting at the Fermi points in the graphene sheet. In the
case of nonzero energy, the up-spin and down-spin states have the
exact high symmetries of spin, forming the pseudospin singlet
pairing. We describe the peculiar and unconventional quantum Hall
effects of the $ n = 0 $ Landau level in monolayer graphene on the
basis of the index theorem and the high degree of symmetry
\cite{park,ezawa}.

This chapter is written as follows. We explain basic properties of
graphene in section II. In section III, we discuss a path integral
of coherent states in brief. In section IV, supersymmetry is
introduced in graphene. In the following section, we investigate the
Atiyah-Singer index theorem and topological properties. The $
\mathrm{G} $ index theorem and deformation is covered. In section
VI, SUSY QM and higher spin symmetry are described. Next we
illuminate the low energy spectrum and unconventional quantum Hall
effects in monolayer graphene. And finally we come to summary and
conclusion.

\section{Basic properties of graphene}

Graphene is a molecule that is composed of carbon atoms placed on a
two dimensional honeycomb lattice. The basic plaquette of the
lattice has a hexagon and the atoms are located at the sites of the
lattice. Electronic properties of graphene can be described by the
tight binding model on which spinless electrons move from site to
site along the links of the lattice without interaction each other.
Under the tight-binding approximation graphene can be expressed by a
simple Hamiltonian of coupled fermions on a hexagonal lattice
\cite{castroneto,gonzalez}. The model Hamiltonian is a form given by
\begin{eqnarray}
\mathrm{H} = -\frac{2 v_{\mathrm{F}}}{3} \sum_{<i,j>}
c^{\dagger}_{i} c_{j},
\end{eqnarray}
where $ <i,j> $ indicates nearest neighbors on the lattice. $
c^{\dagger}_{i} $ and $ c_{i} $ are the creation and annihilation
operators of the fermions located at site $ i $ with anticommuation
relation $ \{ c_{i}, c^{\dagger}_{j} \} = \delta_{ij} $.

In order to calculate the spectrum of Hamiltonian (1), we account
for a periodicity of honeycomb lattice which leads to a Fourier
transformation. The periodic structure provides the energy
eigenvalue problem for the Hamiltonian in a unit cell. The unit cell
consists of two neighboring carbon atoms called $ A $ and $ B $.
They can be expressed by the three vectors $ \vec{u}_{i}, \forall
i=1,2,3$. Under the Fourier transformation of $ c(\vec{p}) =
\sum_{i} e^{ i \vec{p}\cdot \vec{u}_{i} } c_{i} $, the Hamiltonian
is rewritten in terms of
\begin{eqnarray}
\mathrm{H} = -\frac{2 v_{\mathrm{F}}}{3} \int \int d^{2}p \left(
c^{\dagger}_{A}(\vec{p}), c^{\dagger}_{B}(\vec{p}) \right) \left(
\begin{array}{cc} 0 &
\sum_{i=1}^{3} e^{i\vec{p}.\vec{u}_{i}} \\
\sum_{i=1}^{3} e^{-i\vec{p}.\vec{u}_{i}} & 0 \end{array} \right)
\left( \begin{array}{cc} c_{A}(\vec{p}) \\ c_{B}(\vec{p})
\end{array} \right), \end{eqnarray}
where $ c_{A}(\vec{p}) $ and $ c_{B}(\vec{p}) $ denote the Fourier
transformed operators corresponding to the carbon atoms $ A $ and $
B $, respectively.

Now it is easy to take the eigenvalue of the energy for electrons of
graphene. The dispersion energy is given by
\cite{pachos,park,gonzalez}
\begin{eqnarray}
E(p) = \pm \frac{2 v_{\mathrm{F}}}{3} \sqrt{ 1 + 3 \cos^{2}\frac{
\sqrt{3} p_{y} }{2} + 4 \cos\frac{ 3p_{x} }{2} \cos\frac{3p_{y}}{2}
}
\end{eqnarray}
where the lattice distance between atoms becomes normalized to the
unity. From the dispersion relation obtained above, graphene can
have two independent Fermi points, $ \vec{p} = K_{\pm} = \pm
\frac{2\pi}{3}(1,\frac{1}{\sqrt{3}}) $. We can expand it and then
linearize it near the conical singularities of the Fermi points.
Corresponding to the $ K_{+} $ and $ K_{-} $ at the half-filling
case, the Hamiltonian is expressed by the Dirac operators
\begin{eqnarray}
\mathrm{H}^{\pm} = \pm v_{\mathrm{F}} \sum_{\mu=x,y} \gamma^{\mu}
p_{\mu},
\end{eqnarray}
where $ p_{\mu} = -i \hbar \partial_{\mu} $ is the covariant
momentum and the Dirac matrices $ \gamma^{\mu} $ indicate the Pauli
matrices $ \gamma^{\mu} = \sigma^{\mu} $. Hence the low energy
theory of graphene is described by means of free fermions.

The Hamiltonian can be written in the matrix form \cite{pachos,park}
\begin{eqnarray}
\mathrm{H}^{\pm} = \left(  \begin{array}{cc} 0 & D_{\pm} \\
                          D_{\pm} & 0 \end{array} \right).
\end{eqnarray}
Here $ D_{\pm} $ is a Dirac operator given by
\begin{eqnarray}
D_{\pm} = \pm v_{\mathrm{F}} ( \sigma^{x} p_{x} + \sigma^{y} p_{y}
),
= \mp 2i \hbar v_{\mathrm{F}} \left(  \begin{array}{cc} 0 & \partial_{z}  \\
                 \partial_{\bar{z}} & 0 \end{array} \right).
\end{eqnarray}
Here we have expressed $ D_{\pm} $ in terms of $ \partial_{z} =
\frac{1}{2}( \partial_{x} -i \partial_{y} ) $ and $
\partial_{\bar{z}} = \frac{1}{2}( \partial_{x} +i \partial_{y} ). $
On the complex coordinates we can write
\begin{eqnarray}
\nabla^{2} = 4 \partial_{z} \partial_{\bar{z}}.
\end{eqnarray}
So far, we have discussed the collective behavior of graphene
electrons that can be governed by the Dirac equation. In particular,
the velocity of the electrons is effectively 300 times smaller than
the speed of light.

Next let us describe a curved graphene. On the curved surface of
graphene the low energy physics can be illuminated by the Dirac
equation that is defined on the corresponding curved manifold. The
curvature generates an gauge field of magnetic flux going through
the curved graphene. This yields to a picture about how to interact
gauge fields with Dirac fermions. In order to have the way
associated with the gauge fields, let us take into account a good
method to include curvature to graphene. The simplest way is that we
cut a $ \frac{\pi}{3} $ piece of triangle from a graphene sheet and
then gluing the opposite ends of the lattice. This process results
in a single pentagon at the apex of the generated cone while all the
other plaquette keep a hexagon. Due to the minimal geometrical
distortion, the honeycomb lattice has a positive curvature. The
curvature can be obtained by calculating a circular tangent vector,
$ \mathrm{V} $, around the apex by $ \oint \mathrm{V}\cdot d \vec{r}
= \frac{\pi}{3} $. The generation of a single pentagon gives rise to
a dramatic effect on the spinor, resulting in deformation of the
lattice. If the spinor is parallel transported around the apex by an
angle $ 2 \pi $, it is forced at some point to make a jump from a
site $ A $ to a site $ A $ while every site $ A $ takes only $ B $
neighborhoods or vice versa. This motion enables us to have the
effect that the magnetic field gives on the wave function of a
particle moving on a closed path. A full circulation provides
accumulation of all phase factors to the particle wave function, and
so generates the enclosed magnetic flux. This flux allows us to
observe the Aharonov Bohm effect. Quantum mechanically, it can
provide a discontinuity. Thus we have simultaneously to describe it
through the wave function of the particle in the processes of being
static or moving along the closed trajectory. This produces a vector
potential term in the Hamiltonian that leaves the theory to be
consistent.

By similar procedures made on the curved graphene that compensates
the jump in the components of the spinor, we should take into
account a nonabelian vector potential $ \mathrm{A} $ in the
effective Hamiltonian. Around the apex we can take the circulation
of $ \mathrm{A} $ along a path. It is expressed by $ \oint_{\gamma}
\mathrm{A} \cdot d \vec{r} = \frac{\pi}{2} \tau_{2} $, where $
\tau_{2} $ denotes the second Pauli matrix which couples the $ K_{+}
$ to the $ K_{-} $ components of the spinor. The effective gauge
theory can be emerged due to the geometric deformation on geometric
variants of graphene as topological effects. In next section we
describe a path integral of the coherent states for geometrical and
topological properties.

\section{Path integral of coherent states }

Let us construct a path integral of the coherent states on a group
manifold. Following the Stone's approach \cite {stone}, we discuss
the coherent states on a general group $G$. Let us take $D(g)$, any
$g \in G $, as an irreducible representation of the group. Assume
that $ |0> $ is some state in the space of representation. Then one
can define  $ |g> $ by
\begin{equation}
 | g > = D(g) | 0 >.
\end{equation}
On the basis of the irreducible representation, Shur's lemma holds
that
\begin{equation}
 \frac {1}{V(G)} \int d[g] | g ><g | = 1
\end{equation}
where $ \frac {1}{V(G)} $ is the volume of the group manifold. In
the expression of Eq. (9) $ d[g] $ is the Haar measure on the group.

Let us compute a thermodynamic partition function
\begin{equation}
 Z = \mathrm{Tr} ( e^{-\beta \mathrm{H}} )
\end{equation}
where $ \beta $ is the imaginary time. In the procedure of
calculating the partition function, the trace is constrained to the
representation space on which $ D(g)$ acts. Dividing the Matsubara
time-interval $ \beta $ into $n$ parts, and using the Shur's lemma
of Eq. (9), we can write down an iterated integral
\begin{equation}
\begin{array}{ccc}
\mathrm{Tr} ( e^{-\beta \mathrm{H}} ) = \mathrm{const.} \int ( d[g]
d[g^{'}] \cdots ) < g |e^{-\beta \mathrm{H}/n} | g^{'}><g^{'} |
e^{-\beta \mathrm{H}/n} \cdots | g >.
\end{array}
\end{equation}
where $ \mathrm{const.} $ is the constant value taken on the
representation space. When taking into account short time intervals,
one can express $ g^{'} \simeq g + \delta g $, and $ \delta g \simeq
O(\delta t) $, so that
\begin{equation}
\begin{array}{ccc}
< g |e^{-\delta t \mathrm{H}} | g^{'} > \\
\approx 1 + < g | \delta g > + < g |(-\delta t \mathrm{H}) | g > +
O(\delta t^{2}).
\end{array}
\end{equation}
On taking into consideration up to order $ O(\delta t) $, we can
write down the formal path-integral expression
\begin{equation}
\begin{array}{ccc}
\mathrm{Tr} ( e^{-\beta \mathrm{H}} ) &&\\
= \frac {1}{V(G)} \int d[g] \mathrm{exp} ( \oint < g | \delta g > -
\int^{\beta}_{0} dt < g |\mathrm{H} | g > ).
\end{array}
\end{equation}
Here $ d[g] $ is regarded as the path-integral measure given by the
Haar measure at each time step.

The path-integral expression can be identified as a path integration
over a quotient space of the group. In particular it is noted that a
set of the $ | g > $ can be different from only a phase so that the
integrand is not sensitive to the phase factor. Now suppose that $ H
$ is the subgroup of $ G $, constructed from exponentiating a
maximal commuting set of generators, i.e., a maximal torus. Then $ |
0 > $ is expressed by an eigenstate of the generators of $ H $ which
means a state of definite weight. The $ | g > $ are represented by
all phase multiples of one another in any one coset of $ G/H $. And
hence the coherent states can be described in terms of a bundle over
$ G/H $ with the maximal torus as the gauge group, and the
integration is made over the path in $ G/H $.

Let us express the integrand in the coherent-state path integral
without any choice of representatives. In order to make a natural
procedure on the independent choice of the representatives, we
define the projection operators as
\begin{equation}
P(g) = | g ><g |.
\end{equation}
They can be directly projected onto the physically distinct states
since they do not have any phase ambiguity. In the integrand, the
first term $ \oint < g | \delta g > $ can be taken to be a gauge
invariant form by using Stokes theorem
\begin{equation}
\oint_{\Gamma = \partial \Omega} < g | d g > = \int_{\Omega} d < g |
d g >.
\end{equation}
Making use of the identity form
\begin{equation}
d < g | d g > =  < d g | d g > = - \mathrm{Tr} ( d P P d P),
\end{equation}
the first term yields
\begin{equation}
\oint_{\Gamma = \partial \Omega} < g | d g > = - \int_{\Omega}
\mathrm{Tr} ( d P P d P).
\end{equation}
The second term in the integrand can be rewritten by
\begin{equation}
< g | \mathrm{H} | g > =  Tr ( P(g) \mathrm{H}).
\end{equation}
Hence combination of Eqs. (17) and (18) leads us to the path
integral given by \cite{stone}
\begin{equation}
\mathrm{Tr} ( e^{-\beta \mathrm{H}} ) = \frac {1}{V(G/H)} \int d[g]
\mathrm{exp} \left(  - \int_{\Omega} \mathrm{Tr} ( d P P d P) -
\int_{\Gamma} \mathrm{Tr} ( P(g) \mathrm{H}) \right).
\end{equation}
It is noted that the elements of $ G/H $ can only contribute to the
integrand, so that the volume factor of the gauge group $ H $ gets
out of the path integral.

There are certain subgroups corresponded to the Lie algebra $
\mathcal{G} $ of a Lie group $ G $. Let us describe some basic facts
about semisimple Lie algebra in brief. Then it is known that the
generators of the  $ \mathcal{G} $ can have a decomposition into a
maximally commuting set $ \mathcal{H} $ = $ \{ H_{i} \} $, i.e., the
Cartan subalgebra, and a set of ladder operators, $ E_{\alpha}$, one
for each root vector $ \alpha \in R $. The ladder operators are
needed for complexifing the algebra to $ \mathcal{G}^{c} $ which
results from the group parameters to get the complex values. The $
H_{i}$ and the $ E_{\alpha} $ hold that
\begin{equation}
[ H_{i},  E_{\alpha} ] =  \alpha_{i} E_{\alpha}.
\end{equation}
Assume that $ | \lambda > $ is an eigenvector of the $ H_{i}$ with
eigenvalues $ \lambda_{i} $ such as
\begin{equation}
 H_{i} | \lambda > =  \lambda_{i} | \lambda >.
\end{equation}
Then
\begin{equation}
E_{\alpha}  | \lambda > =  | \lambda + \alpha >.
\end{equation}
The roots may be classified into two sets by an arbitrary hyperplane
on a root space. On one set the root objects are positive roots
indicated by $ \alpha \in R_{+} $ as increasing the weights while
the others are negative roots, by $ \alpha \in R_{+} $ as decreasing
the weights. And thus we can take the greatest weight as a state
which is annihilated by all $ E_{\alpha} ,  \alpha \in R_{+} $.

This decomposition of the Lie algebra is made on some kinds of
special subgroup. The Borel subgroups $ B_{\pm} $ are constructed by
exponentiating the algebras $ \mathcal{B_{\pm}} $ which are spanned
by the  $ E_{\alpha} $, $ H_{i}$; $ \alpha \in R_{\pm} $, $ H_{i}
\in \mathcal{H}^{c} $. Any $ g \in G $ can be decomposed into the
gaussian factors
\begin{equation}
\begin{array}{ccc}
g = \zeta_{-} h \zeta_{+}; &&  \zeta_{-} \in Z_{-},  \zeta_{+} \in
Z_{+},  h \in H_{+}^{c},
\end{array}
\end{equation}
where the $ Z_{\pm} $ are the groups obtained by exponentiating the
$ E_{\alpha} ,  \alpha \in R_{\pm} $. As an example, the gaussian
factor $ \zeta_{-} $ can be written as
\begin{equation}
\zeta_{-} = \mathrm{exp} ( \sum_{ \alpha \in R_{-}} z^{\alpha}
E_{\alpha} ).
\end{equation}
When the representation matrix $ D(g) $ is applied to a greatest
weight, the  factor $ \zeta_{+} $ is regarded as the identity. The $
z^{\alpha} $ can be taken as complex coordinates for the coset space
$ G^{c}/B_{+} $. It follows that the set of physically distinct
states can have an one-to-one correspondence to the $ G^{c}/B_{+} $.
For a complex manifold we can choose complex coordinates on the
manifold with holomorphic functions. And constructed on non-greatest
weight states, coherent states are described by nonholomorphic
functions of $ \bar {z} $. In the next section, after introducing
supersymmetry on the complex manifold in brief, we will describe
supersymmetric quantum mechanics in graphene \cite{witten,park}.

\section{Supersymmetry in graphene}

Let us use the methods of Witten to introduce fermionic creation
operators $ \psi^{x\dag} $ and $ \psi^{y\dag} $ which correspond to
the differential forms $ \mathrm{d}x $ and $ \mathrm{d}y $
\cite{witten,park}
\begin{eqnarray}
\mathrm{d}x \leftrightarrow  \psi^{x\dag} |0>, & \mathrm{d}y
\leftrightarrow  \psi^{y\dag} |0> .
\end{eqnarray}
In more detail $ \psi^{x\dag} $ performs the operation of exterior
multiplication by $ \mathrm{d}x $ while the adjoint, $ \psi_{x} $,
does that of interior multiplication by the vector dual to $
\mathrm{d}x $, say, $ \partial_{x} $
\begin{eqnarray}
\imath : \partial_{x} \rightarrow  \psi_{x}, & \mathcal{E} :
\mathrm{d}x \rightarrow  \psi^{x\dag}, \quad \imath :
\partial_{y} \rightarrow  \psi_{y}, & \mathcal{E} : \mathrm{d}y
\rightarrow  \psi^{y\dag}.
\end{eqnarray}
The fermionic operators satisfy the anticommutation relations
 \begin{eqnarray}
\{ \psi_{\mu}, \psi^{\nu\dag} \} = \delta_{\mu}^{\nu}, & \mu, \nu =
x, y.
\end{eqnarray}
On the basis of these definitions let us set up
\begin{eqnarray}
\frac{1}{2} ( \psi_{x} - i\psi_{y} ) = \psi_{z} =
\frac{1}{2}\psi^{\bar{z}}, \quad \frac{1}{2} ( \psi_{x} + i\psi_{y}
) = \psi_{\bar{z}} = \frac{1}{2}\psi^{z}.
\end{eqnarray}
while holding the Hermitian conjugate relations.

Now let us describe a supersymmetry over the complex K$\ddot{a}$hler
manifold. On the complex manifold we take two supercharges
\begin{eqnarray}
Q_{1} = \partial = \psi^{\dag z} \partial_{z}, &
Q_{1}^{\dag} = \delta = -\psi^{\bar{z}} \partial_{\bar{z}}, \nonumber\\
Q_{2} = \bar{\partial} = \psi^{\dag \bar{z}} \partial_{\bar{z}}, &
Q_{2}^{\dag} = \bar{\delta} = -\psi^{z} \partial_{z}.
\end{eqnarray}
Making use of anticommutation relations for fermions, these
supercharge operators allows us to express
\begin{eqnarray}
\partial \delta + \delta \partial = Q_{1} Q_{1}^{\dag} + Q_{1}^{\dag} Q_{1}
= -\frac{1}{2} \nabla^{2}, \nonumber\\
\bar{\partial} \bar{\delta} + \bar{\delta} \bar{\partial} = Q_{2}
Q_{2}^{\dag} + Q_{2}^{\dag} Q_{2} = -\frac{1}{2} \nabla^{2}.
\end{eqnarray}
Here it is easy to check that the cross terms such as
\begin{eqnarray}
\partial \bar{\delta} + \bar{\delta} \partial = Q_{1} Q_{2}^{\dag} + Q_{2}^{\dag} Q_{1}
= 0,
\end{eqnarray}
do not have any contributions.

In terms of the two supercharge operators, the Dirac operators $
D_{\pm} $ in graphene is given by
\begin{eqnarray}
D_{\pm} = \mp i\hbar v_{\mathrm{F}} ( Q_{1} + Q_{2} ), \equiv
Q_{\pm}.
\end{eqnarray}
The Dirac operators can be described by means of the sum of the two
supercharges which is the ordinary exterior derivative, $ \mathrm{d}
$. And $ Q_{\pm} $ become the supercharge of $ \mathrm{ N } $ = 1
supersymmetric quantum mechanics. It is not hard to check up that $
D_{\pm} = Q_{\pm} + Q_{\pm}^{\dag} $ is given by
\begin{eqnarray}
Q_{\pm} + Q_{\pm}^{\dag} =  \mp 2i \hbar v_{\mathrm{F}} \left(
                \begin{array}{cc} 0 & \partial_{z}  \\
                 \partial_{\bar{z}}  & 0 \end{array} \right)
\end{eqnarray}
which are equivalent to Eq. (6). Furthermore the square of Eq. (33)
can lead to
 \begin{eqnarray}
(Q_{\pm} + Q_{\pm}^{\dag})^{2} &=& - 4(\hbar v_{\mathrm{F}})^{2}
\left(
\begin{array}{cc}  \partial_{z} \partial_{\bar{z}}  &  0 \\
                 0 & \partial_{\bar{z}} \partial_{z} \end{array} \right)
= -\hbar^{2} v_{\mathrm{F}}^{2} \nabla^{2},
\end{eqnarray}
where we have exploited Eq. (7). And hence in the sense of SUSY QM
the Hamiltonian of graphene may be recapped in terms of the
following form
\begin{eqnarray}
\mathrm{H} \equiv 2 (Q_{\pm} + Q_{\pm}^{\dag})^{2} &=& -2 \hbar^{2}
v_{\mathrm{F}}^{2} \nabla^{2}, \equiv - \frac{ \hbar^{2} \nabla^{2}
}{2 m^{*}},
\end{eqnarray}
provided that the mass, $ m^{*} $ were defined by $ m^{*} =
\frac{1}{ 4 v_{\mathrm{F}}^{2} } $ in the last expression of the eq.
(17). The Witten index is given by
\begin{eqnarray}
\mathrm{Index} ( \mathrm{d} ) = \mathrm{Tr} \left( (-1)^{F} e^{ -t
\mathrm{H} } \right)
\end{eqnarray}
which accounts for the Euler number of the manifold as the exterior
calculus of the de-Rham complex. In order to build up the
supersymmety over the Dolbeault complex \cite{eguchi}, we need one
of the supercharges, $ Q_{2} = \bar{\partial} $. On the manifold of
real dimension $ 2n $, the index of the Dolbeault complex is given
by
\begin{eqnarray}
\mathrm{Index} ( \bar{\partial} ) = \mathrm{Tr} \left( (-1)^{F} e^{
-t (Q_{2} + Q_{2}^{\dag})^{2} } \right)
\end{eqnarray}
This index is more interested in the the SUSY QM over twisted
Dolbeault complex which is associated with deformation of the
topology of the lattice on a graphene system . And in the next
section we build the Atiyah-Singer and G index theorem as well as
the energy eigenvalues on the deformation of the compact manifold.

\section{\label{sec:level1} Index theorem in graphene}

Let us describe the index theorem which gives an insight on the
spectrum structure of certain operators such as the Dirac operators.
In graphene this theorem enables us to have physical properties
associated with the topology and geometry of the space in which the
Dirac operators are defined. It provides the relationship between
the analytic properties of the operator and the topological
characteristics of the manifold.

\subsection{The Atiyah-Singer index theorem in graphene}

We illuminate the Atiyah-Singer index theorem by the method employed
to the heat kernel expansion. The theorem furnishes a relation
between zero eigenvalues of the Dirac operator of graphene and the
total flux which goes through its surface. If the latter is
connected to the genus of the surface through the Euler
characteristic, we can find a close relation between the zero modes
and the topology of the surface on a graphene system.

Let us start with a Dirac operator given by
\begin{eqnarray}
\mathrm{K} =  \left(
\begin{array}{cc} 0 & D^{\dagger} \\
D & 0 \end{array} \right).
\end{eqnarray}
Here $ D $ means an operator that maps a space $ M_{+} $ onto a
space $ M_{-} $ while $ D^{\dagger} $ is a map from $ M_{-} $ to $
M_{+} $. If $ D $ is an $ n \times m $ matrix, $ D^{\dagger} $
becomes a $ m \times n $ matrix. $ M_{+} $ and $ M_{-} $ are the
space of $ n $ and $ m $ dimensional vectors, respectively. Because
we are focusing on the zero modes of $ \mathrm{K} $ such that the
solutions of the equations $ \mathrm{K} \Psi = 0 $, let us define
the number of different eigenstates of $ D $ with zero eigenvalue as
$ \eta_{+} $ and the ones of $ D^{\dagger} $ as $ \eta_{-} $. As a
bookkeeping of chirality, the chirality operator $ \gamma_{5} $ is
defined as
\begin{eqnarray}
\gamma_{5} =  \left(
\begin{array}{cc} 1 & 0 \\
0 & -1 \end{array} \right).
\end{eqnarray}
Its eigenstates can have eigenvalue $ \pm $ provided that they act
on $ M_{\pm} $.

In order to calculate the number of zero eigenstates in which we are
interested, we take into account the operator $ \mathrm{K}^{2} $
that has the same number of zero modes as $ \mathrm{K} $. The $
\mathrm{K}^{2} $ can be given by a diagonal form
\begin{eqnarray}
\mathrm{K}^{2} =  \left(
\begin{array}{cc} D^{\dagger} D & 0 \\
0 & DD^{\dagger} \end{array} \right).
\end{eqnarray}
It is claimed that the operators $ D^{\dagger} D $ and $
DD^{\dagger} $ get the same non-zero eigenvalues. To prove this
statement, assume that $ DD^{\dagger} \psi = \lambda \psi $ for
eigenvalue $ \lambda \neq 0 $. Then it follows that
\begin{eqnarray}
DD^{\dagger} \psi = \lambda \psi \longrightarrow
D^{\dagger}D(D^{\dagger} \psi) = \lambda (D^{\dagger}\psi).
\end{eqnarray}
This means that the operator $ D^{\dagger}D $ gives the same
eigenvalue, $\lambda$, which corresponds to the eigenstate
$D^{\dagger}\psi$. But it is not necessary to hold the case for
$\lambda = 0$ when $D^{\dagger}\psi$ might be zero by itself.

Let us compute the trace of $ \gamma_{5} e^{-t \mathrm{K}^{2} } $ as
followings:
\begin{eqnarray}
\mathrm{Tr} (\gamma_{5} e^{-t \mathrm{K}^{2} }) = \mathrm{Tr} (e^{-t
D^{\dagger}D }) - \mathrm{Tr} (e^{-tDD^{\dagger} }) =
\sum_{\lambda_{+}} e^{-t \lambda_{+}} - \sum_{\lambda_{-}} e^{-t
\lambda_{-}},
\end{eqnarray}
where $ \lambda_{+} $ and $ \lambda_{-} $ indicate the eigenvalues
of the operators $ D^{\dagger}D $ and $ DD^{\dagger} $,
respectively, and $ t $ is an arbitrary parameter. In the first step
of the above procedures, $ \gamma_{5} $ acts on the exponential so
that it provides a $ + 1 $ to the eigenvectors of $ D^{\dagger}D $
when they are placed in $ M_{+} $, and a $ -1 $ to the ones of $
DD^{\dagger} $ when they belong to $ M_{-} $. In the last step the
trace is evaluated by a sum over all the eigenvalues of the
corresponding operators. Every non-zero eigenvalue of $ D^{\dagger}D
$ is a one-to-one correspondence to an eigenvalue of $ DD^{\dagger}
$. Therefore all paired terms of non-zero eigenvalues cancel out
each other. There are left over the zero eigenvalues of each
operators, resulting in
\begin{eqnarray}
\mathrm{Tr} (\gamma_{5} e^{-t \mathrm{K}^{2} }) = \eta_{+} -
\eta_{-}.
\end{eqnarray}
In general we cannot determine difference between the number of zero
modes. It is seen that the above result is independent of t owing to
the cancelation of the non-zero eigenvalue term.

Actually we should evaluate $ \mathrm{Index}(\mathrm{K}) $. In order
to calculate it practically, we take an alternative method of heat
expansion for calculating $ \mathrm{Tr} (\gamma_{5} e^{-t
\mathrm{K}^{2} }) $. It says that for general $ \hat{\Gamma} $ and $
\hat{D} $ on a two dimensional compact manifold we can expand
\begin{eqnarray}
\mathrm{Tr} (\hat{\Gamma} e^{-t \hat{D} }) = \frac{1}{4\pi t}
\sum_{l \ge 0} t^{\frac{l}{2}} b_{l}(\hat{\Gamma},\hat{D}),
\end{eqnarray}
where $ \mathrm{Tr} $ indicates the trace of matrices and the
integration over coordinates of space. $ b_{l} $ denote expansion
coefficients. For $ \hat{\Gamma} = \gamma_{5} $ and $ \hat{D} =
\mathrm{K}^{2} $, we have to return to an expression that is
t-independent. For this t-independence, the expansion coefficients
should vanish for all $ l $ except for $ l = 2 $ under the condition
that all the $ t $ contributions is canceled out each other. This
allows us to determine the coefficient $ b_{2} $ from the first
order term in $ t $ in the series of expansion. For the evaluation,
let us take $ D $ as $ D = -ie_{\nu}^{\mu} \sigma^{\nu}(
\nabla_{\mu} -ieA_{\mu} ) $. Here $ e_{\nu}^{\mu} $ indicates the
zweibein of curved surface metric $ g_{\mu \nu} $ that defines a
local flat frame $ \eta_{\alpha \beta} = e_{\alpha}^{\mu}
e_{\beta}^{\nu} g_{\mu \nu} $ while $ \sigma_{\mu} $ is the Pauli
matrix. $ A_{\mu} $ denotes a gauge field. It follows that
\begin{eqnarray}
\mathrm{K}^{2} = -g^{\mu \nu} \nabla_{\mu} \nabla_{\nu} +
\frac{1}{4} [\gamma^{\mu},\gamma^{\nu}]F_{\mu \nu} - \frac{1}{4}R
\end{eqnarray}
where $ R $ is curvature, and $ F_{\mu \nu} =
\partial_{\mu}A_{\nu} - \partial_{\nu}A_{\mu} $ means the
field strength. $ \nabla_{\mu} $ indicates a covariant derivative
with respect to gauge and reparametrization transformation.

It is easy to see that the non-zero expansion coefficient $ b_{2} $
is written by
\begin{eqnarray}
b_{2} = \mathrm{Tr}[ \gamma_{5}(\frac{i}{4}[
\gamma^{\mu},\gamma^{\nu}]F_{\mu\nu} - \frac{1}{4}R ) ] = 2\int\int
\mathrm{B}\cdot d \mathrm{S},
\end{eqnarray}
where $ \mathrm{B} $ is the magnetic field given by $ B_{l} =
\frac{1}{2} \epsilon^{l\mu\nu} F_{\mu\nu} $. The integration has
been over the whole surface. These two independent ways for
calculating $ \mathrm{Tr} (\gamma_{5} e^{-t \mathrm{K}^{2} }) $
allows u to arrive at the final formula of the index theorem
\cite{pachos}
\begin{eqnarray}
\mathrm{Index}(\mathrm{K}) = \nu_{+} - \nu_{-} = \frac{1}{2\pi}
\int\int \mathrm{B} \cdot d \mathrm{S}.
\end{eqnarray}
It states that the total flux which goes out of the surface is
related to the number of zero modes of the $ \mathrm{K} $ operator.
The curvature doesn't contribute to the index formula because $
\gamma_{5} $ is a traceless operator. Particularly no contribution
of curvature is to show an intrinsic property of two dimensional
surfaces. The index theorem leads to an integer number on the
compact surfaces. Therefore $ \int \int \mathrm{B} \cdot d
\mathrm{S} $ produces the total magnetic monopole charge in discrete
values inside the surface under the Dirac quantization condition of
magnetic monopoles.

Using the index theorem, we observe the topological characteristics
of graphene. Let us apply it to graphene and its geometric variants.
In a certain configuration of graphene we have to account for the
effective magnetic field on the graphene surface. Some plaquette
deformations give rise to a specific circulation of the vector
potential around a loop trajectory. Stokes's theorem helps us to
have a relation between the circulation of the gauge potential
around a loop $ \gamma_{i} $ and the flux of the corresponding
magnetic field
\begin{eqnarray}
\oint_{\gamma_{i}} \mathrm{A}\cdot d\vec{r} = \int\int_{S_{i}}
\mathrm{B} \cdot d \mathrm{S},
\end{eqnarray}
where $ S_{i} $ indicates the area. Hence going through the surface
of graphene the total flux can be obtained from the fact that we can
know the total number of deformations. It is necessary for us to
have some information about the total number of plaguette
deformations. The information is related to the topological
properties of the surface through the Euler characteristics. Let us
consider the Euler theorem in the following subsection.

\subsection{Euler theorem }

In general the Euler theorem gives rise to a relation between the
structural information of a polyhedral lattice and its topological
properties. There are a lot of proofs on this theorem. The most
common methods are proofs on the basis of a reductions from the
polyhedral lattice to the simpler one without changing its
topological properties.

Let us consider a lattice placing on a compact surface with a
certain genus $ g $. Then we can compute the number of deformations
in a lattice necessary to create such a surface by applying the
Euler characteristic. Let $ V, E $ and $ F $ be the number of
vertices, edges and faces of the lattice, respectively, and $
N_{end} $, open ends. Then the Euler characteristic, $ \chi $ is
expressed by \cite{pachos}
\begin{eqnarray}
\chi = V - E + F = 2(1 -g) - N_{end}.
\end{eqnarray}
The second step of Eq. (49) is satisfied by the Euler theorem. It is
easily to check that a single cut in the surface can have a
reduction of the genus by one and increase the number of open ends
by two, say, $ (g,N_{end}) \rightarrow (g-1,N_{end}+2) $, leaving
the Euler characteristic $ \chi $ preserved.

Let us apply the Euler theorem to the case of graphene molecules.
There are three links on each vertex of graphene. Suppose that
topological deformations such as pentagons or heptagons are present.
Let us indicate the total number of pentagons, hexagons and
heptagons by $ n_{5}, n_{6} $ and $ n_{7} $, respectively in the
molecule. Then the total number of vertices is written by $ V =
(5n_{5} + 6n_{6} + 7n_{7})/3 $ when each $k$-gon has $ k $ vertices
and each vertex takes three polygons. Similarly, the total number of
edges is expressed by $ E = (5n_{5} + 6n_{6} + 7n_{7})/2 $ provided
that each edge has two polygons. The total number of faces is equal
to the sum of different polygons, $ F = n_{5} + n_{6} + n_{7} $.
Combining these into the Euler characteristic, we see that
\begin{eqnarray}
n_{5} - n_{7} = 6\chi = 12(1 - g) - 6N_{end}.
\end{eqnarray}
This result reflects many facts. When equal numbers of pentagons and
heptagons are inserted, they do not make any change about the
topology of the surface in the case that they cancel out. On a flat
graphene sheet we can put two pentagons and two heptagons on it
without changing the curvature of the molecule away from these
deformations. This is consistent with the effective gauge flux
approaches where pentagons and heptagons give opposite flux
contributions. On the other hand, it is known that nontrivial
topologies necessarily provide an imbalance between pentagons and
heptagons. The genus zero configurations result in an excess of
pentagons while high genus surface has an excess of heptagons. Genus
one surfaces do not have any pentagons or heptagons at all provided
that they are equivalent to a flat sheet.

It is seen that Eq. (50) recaps the known result of a sphere with $
g=0 $ which leads to $ \chi = 2 $. This corresponds to a fact about
$ n_{5} = 12 $ and $ n_{7} = 0 $ for the $ \mathrm{C}_{60} $
fullerene. For a torus with $ g = 1 $, we can have $ \chi = 0 $
related to $ n_{5} = n_{7} = 0 $ in the case of the nanotubes. Thus
no pentagons or heptagons may be required. If we account for the
genus $ g = 2 $ surfaces, then we can obtain $ \chi = -2 $ where $
n_{5} = 0 $ and $ n_{7} = 12 $. In this situation equal numbers of
pentagons and heptagons can be inserted without making any change of
topology on the surface.

Suppose that $ \mathrm{K} $ is a Dirac Hamiltonian $ \mathrm{H} $.
Then we compute the $ \mathrm{Index}(\mathrm{H}) $. The Euler
characteristic term allows us to compute the gauge field term in the
$ \mathrm{Index}(\mathrm{H}) $. It can be given by including
additionally the contributions from the surplus of pentagons or
heptagons. Hence the total flux of the effective gauge field can
lead to
\begin{eqnarray}
\mathrm{Index}(\mathrm{H}) &=& \frac{1}{2\pi} \int\int_{S_{i}}
\mathrm{B} \cdot d \mathrm{S} = \frac{1}{2\pi} \sum_{n_{5}-n_{7}}
\oint_{\gamma_{i}} \mathrm{A} \cdot d\vec{r} \nonumber\\
&=& \frac{1}{2\pi} \frac{\pi}{2}(n_{5} - n_{7}) = 3(1 - g) -
\frac{3}{2}N_{end}.
\end{eqnarray}
The total number of zero modes is equivalent to the sum contributed
from each subsector of a Dirac operator. As a consequence, by adding
the two contributions, we arrive at the index of the Dirac
Hamiltonian that describes the graphene molecule. The $
\mathrm{Index}(\mathrm{H}) $ is expressed by \cite{pachos}
\begin{eqnarray}
\mathrm{Index}(\mathrm{H}) = \nu_{+} - \nu_{-} = 6(1 - g) - 3N_{end}
\end{eqnarray}
which is consistent with the exact number of the zero modes if $
\nu_{+} = 0 $ or $ \nu_{-} = 0 $. Therefore we have obtained the
theorem which relates the number of zero modes existed in a certain
graphene molecule to the topological characteristics of its surface.

This result provides the number of zero modes for the familiar cases
of graphene molecules. For example, since a fullerene takes $ g = 0
$ and $ N_{end} = 0 $, it is expected that it has six zero modes
which correspond to the two triplets of $ \mathrm{C}_{60} $ and of
similar large molecules. For the case of nanotubes, we have $ g = 0
$ and $ N_{end} = 0 $. This results in $ \nu_{+} - \nu_{-} = 0 $
which is in agreement with previous theoretical and experimental
results \cite{reich,saito}. The index theorem gives rise to a
surprising relation between the topology and the presence of
magnetic flux which is effectively inserted in graphene molecules by
geometrical deformations. The number of these deformations can be
associated with the general topological characteristics of the
lattice surface. They are related to the zero modes of a general
graphene molecule with the genus and the number of open faces of its
surface.

\subsection{G index theorem }

To build up supersymmetry over the Dolbeault complex which are
associated with deformation on topology of the lattice
\cite{eguchi}, we need one of the supercharges, $ Q_{2} =
\bar{\partial} $. The index theorem leads to a topological invariant
under deformations on $Q_2$ and plays an essential role in
formulating the SUSY QM over the twisted Dolbeault complex caused by
the deformation in a graphene system.

Let us describe the general statement of the G-index theorem.
Suppose that $ Q $ and $ Q^{\dag} $ are supercharges which have a
map from a space of bosonic states to a space of fermionic states
and vice versa. Further let us take a Lie group $ \mathrm{G} $
generated by $ G_{i} $. $ G_{i} $ satisfy the commutation relations
with $ Q $ and $ Q^{\dag} $ such that \cite{stone}
\begin{eqnarray}
[ G_{i}, G_{j} ] = i f^{k}_{ij} G_{k}, \quad [ Q, G_{i} ]& = 0, [
Q^{\dag}, G_{i} ] = 0,
\end{eqnarray}
$ f^{k}_{ij} $ are structure constants. $ G_{i} $ has also a
commutation relation with the Hamiltonian of Eq. (5)
 \begin{eqnarray}
[ G_{i}, \mathrm{H} ] = 0.
\end{eqnarray}
A supercharacter can be expressed as a supertrace of the group
elements
 \begin{eqnarray}
\Xi ( e^{ i \theta^{i} G{i} } ) =  \mathrm{Tr} \left( (-1)^{F} e^{ i
\theta^{i} G{i} } e^{ -t \mathrm{H} } \right).
\end{eqnarray}
The ordinary trace requires us to take a limit of $ t \rightarrow 0
$. But Eq. (55) does not have any dependence on $ t $ because the
non-zero energy levels can not make contributions due to canceling
in pairs between the bosonic and fermionic sectors. But the
zero-energy levels can only give contributions. And these are not
dependent on $ t $.

The character is topologically invariant under deformations of the
operators $ Q $ and $ Q^{\dag} $. On deformation let us consider the
supersymmetry generator $ Q $ of the ordinary $ \mathrm{N} = 1 $
SUSY QM on a manifold. In locally geodesic coordinates $ Q $ is
expressed by
\begin{eqnarray}
Q = \mathrm{d} = \psi^{\mu \dag}\partial_{\mu}.
\end{eqnarray}
When deforming $ Q $, it is changed into the new operator
\begin{eqnarray}
Q_{s} = \mathrm{d} + s \iota_{\mathcal {K}} = \psi^{\mu
\dag}\partial_{\mu} + s \mathcal {K}^{\mu} \psi_{\mu},
\end{eqnarray}
where $ \mathcal {K} $ denotes a Killing vector field. And then the
deformed Hamiltonian is given by
\begin{eqnarray}
\mathrm{H}_{s}& \equiv& 2 (Q_{s} + Q_{s}^{\dag})^{2},\nonumber\\
&=& -\partial^{2} + s^{2}|\mathcal{K}|^{2} - \frac{1}{2} s [
\psi^{\dag}_{\mu}, \psi^{\dag}_{\nu} ]\partial_{\mu}
\mathcal{K}_{\nu} - \frac{1}{2} s [ \psi_{\mu}, \psi_{\nu}
]\partial_{\nu} \mathcal{K}_{\mu}.
\end{eqnarray}

Now suppose that we decompose the deformed $ N = 1 $ SUSY operator
into holomorphic and antiholomorphic sectors because of $ \mathrm{d}
= \partial + \bar{\partial} $. Then the G-index theorem can be
associated with a high degree of symmetry on graphene molecule. Let
us take a Lie group $ \mathrm{G} $ generated by $ G_{i} $ which
satisfies the commutation relations with $ Q_{2} $ and $
Q_{2}^{\dag} $ such that
\begin{eqnarray}
[ G_{i}, G_{j} ] = i f^{k}_{ij} G_{k},\quad [ Q_{2}, G_{i} ] = 0, [
Q_{2}^{\dag}, G_{i} ] = 0
\end{eqnarray}
where $ f^{k}_{ij} $ mean structure constants. On graphene, there
exists the Hamiltonian $ \mathrm{H} $ such that
 \begin{eqnarray}
\mathrm{H} \equiv - 2\hbar^{2} v_{{\rm F}}^{2} (Q_{2} +
Q_{2}^{\dag})^{2}, \quad [ G_{i}, \mathrm{H} ] = 0.
\end{eqnarray}

On deforming $ Q_{2} $, this can be changed into a new operator
\begin{eqnarray}
Q_{2s} = \psi^{\dag \bar{z}} \partial_{\bar{z}} + s \mathcal
{K}^{\bar{z}} \psi_{\bar{z}},
\end{eqnarray}
where $ s $ is a real parameter. $ \mathcal {K} $ denotes a Killing
vector field associated with the deformation. In the complex
coordinates $z$ and $\bar z$, the Killing vector field is given by $
\mathcal {K} $ = $ -y \partial_{x} + x \partial_{y} = i( z
\partial_{z} - \bar{z} \partial_{\bar{z}} ) $.

Under the deformation let us consider a graphene system
corresponding to the $ K_{+} $. Then we can obtain a deformed
Hamiltonian $\mathrm{H}_{+,s}$ on the Dolbeault complex. For
convenience, after replacing $ s $ by $ is $, the deformed
Hamiltonian is given by \cite{park,stone}
\begin{eqnarray}
\mathrm{H}_{+,is} &=& - 2\hbar^{2} v_{{\rm F}}^{2}(Q_{+,2is} + Q^{\dag}_{+,2is})^{2} \nonumber\\
&=& - 2\hbar^{2} v_{{\rm F}}^{2}\left( -2 \partial^{2}_{\bar{z}z} +
|s|^{2} |z|^{2} \right) - 2\hbar^{2} v_{{\rm F}}^{2} \left( s[
\psi^{\dag \bar{z}}, \psi_{\bar{z}} ] + s (
\bar{z}\partial_{\bar{z}} - z\partial_{z} ) \right).
\end{eqnarray}
This operator gives rise to eigenvalues \cite{park,stone}
\begin{eqnarray}
E_{+,nml} = - 2\hbar^{2} v_{{\rm F}}^{2} \left[ |s| \left( ( n +
\frac{1}{2} ) + ( m + \frac{1}{2} \right) + sl + s( \pm 1) \right].
\end{eqnarray}
Depending on the choice of $ s > 0 $ or $ s < 0 $, we can consider
two cases for zero eigenvalues satisfying the topological invariance
imposed by the index theorem. First if $ s > 0 $, we should take $
-1 $ for the $ [ \psi^{\dag \bar{z}}, \psi_{\bar{z}} ] $. Now we get
in the bosonic sector. We can have zero eigenvalues in the case of $
l = 0, -1, -2, \cdots $. Second if $ s < 0 $, we have to choose $ +
1 $ for $ [ \psi^{\dag \bar{z}}, \psi_{\bar{z}} ] $. And then we can
get zero eigenvalues for $ l = 0, 1, 2, \cdots $.

According to the choice of $ s $, let us take two Hamiltonians $
\mathrm{H}_{+,is}^{\uparrow} = \widehat{Q}^{\dag}_{+,2is}
\widehat{Q}_{+,2is} $ and $ \mathrm{H}_{+,is}^{\downarrow} =
\widehat{Q}_{+,2is} \widehat{Q}^{\dag}_{+,2is}$ as two superpartners
which imply the $Z_2$ grading over the Hilbert space. Here we have
expressed $ \widehat{Q}_{+,2is} = -i\sqrt{2} \hbar v_{{\rm
F}}(Q_{+,2is} + Q^{\dag}_{+,2is}). $ Then we want to calculate the
eigenvalues for the up-spin and down-spin components
\begin{eqnarray}
\mathrm{H}_{+,is}^{\uparrow \downarrow} | \psi_{+,nml}^{\uparrow
\downarrow} > = E_{+,nml}^{\uparrow \downarrow} |
\psi_{+,nml}^{\uparrow \downarrow} >,
\end{eqnarray}
where $ E_{+,n+1ml}^{\uparrow \downarrow} > E_{+,nml}^{\uparrow
\downarrow} \ge E_{+,000}^{\uparrow \downarrow} $. Suppose that $
E_{+,000}^{\uparrow } $ is zero. Then making use of the relations
\begin{eqnarray}
 \widehat{Q}_{+,i2s} \mathrm{H}_{+,is}^{\uparrow} =
 \widehat{Q}_{+,2is} \widehat{Q}^{\dag}_{+,2is} \widehat{Q}_{+,2is}
 = \mathrm{H}_{+,is}^{\downarrow} \widehat{Q}_{+,2is},
\end{eqnarray}
we can have
 \begin{eqnarray}
&\mathrm{H}_{+,is}^{\downarrow} \widehat{Q}_{+,2is} |
\psi_{+,nml}^{\uparrow } > & = \widehat{Q}_{+,2is}
\mathrm{H}_{+,is}^{\uparrow} | \psi_{+,nml}^{\uparrow } > { }
                                                             \nonumber\\
= & E_{+,nml}^{\uparrow } \widehat{Q}_{+,2is} |
\psi_{+,nml}^{\uparrow } >.&
 \end{eqnarray}
 This means that if $ E_{+,nml}^{\uparrow } \neq 0,
 \widehat{Q}_{+,2is} | \psi_{+,nml}^{\uparrow } > $ is an eigenstate of
 $ \mathrm{H}_{+,is}^{\downarrow}. $
 And similarly
\begin{eqnarray}
 &\mathrm{H}_{+,is}^{\uparrow} \widehat{Q}^{\dag}_{+,2is} | \psi_{+,nml}^{\downarrow } > &
= \widehat{Q}^{\dag}_{+,2is} \mathrm{H}_{+,is}^{\downarrow} |
\psi_{+,nml}^{\downarrow } > { }
 \nonumber\\
 = & E_{+,nml}^{\downarrow } \widehat{Q}^{\dag}_{+,2is} | \psi_{+,nml}^{\downarrow } >. &
\end{eqnarray}
If $ E_{+,nml}^{\downarrow } \neq 0, \widehat{Q}^{\dag}_{+,2is} |
\psi_{+,nml}^{\downarrow } > $ is an eigenstate of $
\mathrm{H}_{+,is}^{\uparrow} $. Therefore, for non-zero eigenvalues
there exist the up-spin and down-spin eigenstates in pair. They form
a supermultiplet connected by the supercharge $ \widehat{Q}_{+,2is}
$. The up-spin (down-spin) sector may be described as the bosonic
(fermionic) sector at the $ K_{+} $ point.

In the case of the zero eigenvalues for the down-spin sector, we
should investigate both $ E_{+,000}^{\downarrow} = 0 $ and $
E_{+,000}^{\downarrow}\neq 0 $ separately. Let us assume that $
E_{+,000}^{\downarrow}\neq 0 $. Then the lowest eigenstate of $
\mathrm{H}_{+,is}^{\downarrow} $ is expressed as $ |
\psi_{+,000}^{\downarrow } > $ $ \propto $ $ \widehat{Q}_{+,2is} |
\psi_{+,1ml}^{\uparrow } >. $ Hence, there exists a supermultiplet
between the states $ | \psi_{+,n+1ml}^{\uparrow } > $ and $ |
\psi_{+,nml}^{\downarrow } > $, which have the same energy $
E_{+,n+1ml}^{\uparrow } = E_{+,nml}^{\downarrow } $, and we obtain
energy eigenstates \cite{park}
\begin{eqnarray}
| \psi_{+,nml}^{\downarrow } > &=& \frac{1}{\sqrt{
|E_{+,n+1ml}^{\uparrow }| } }
\widehat{Q}_{+,2is} | \psi_{+,n+1ml}^{\uparrow } >, \nonumber\\
| \psi_{+,n+1ml}^{\uparrow } > &=& \frac{1}{\sqrt{
|E_{+,nml}^{\downarrow }| } } \widehat{Q}_{+,2is}^{\dag} |
\psi_{+,nml}^{\downarrow } >
\end{eqnarray}
for $ n \geq 0. $ If $ E_{+,000}^{\downarrow} = 0 $ and $
\widehat{Q}^{\dag}_{+,2is}  | \psi_{+,000}^{\uparrow } > = 0, $ the
relationships are written by \cite{park}
\begin{eqnarray}
| \psi_{+,nml}^{\downarrow } > &=& \frac{1}{\sqrt{
|E_{+,nml}^{\uparrow }| } }
\widehat{Q}_{+,2is} | \psi_{+,nml}^{\uparrow } >, \nonumber\\
| \psi_{+,nml}^{\uparrow } > &=& \frac{1}{\sqrt{
|E_{+,nml}^{\downarrow }| } } \widehat{Q}_{+,2is}^{\dag} |
\psi_{+,nml}^{\downarrow } >
\end{eqnarray}
for $ n \geq 1. $ Similarly, one can repeat eigenvalue problem for $
\mathrm{H}_{-,is} $, which corresponds to the $ K^{'}_{-} $ point.
As a relation between the up-spin and down-spin eigenstates, the
bosonic (fermionic) sector is regarded as the down-spin (up-spin)
sector at the $ K^{'}_{-}$ point. And hence there exists the 4-fold
degenerate energy spectrum.

\section{\label{sec:level1} Deformed energy eigenvalues \\
and unconventional quantum Hall effect}

On a sheet of graphene, let us consider the problem of magnetic
field concentrated on a thin cylindrical shell of small, but finite
radius $ l_{B} = \sqrt{ \frac{c\hbar}{eB} } $. The corresponding
vector potential is given by $\vec{a} = (-y, x)/2 l^{2}_{B} $ on the
two dimensional plane of graphene. The problem in question is to
compute the eigenvalues of the Dirac Hamiltonian in the field of a
fractional magnetic flux on the graphene sheet
\cite{jackiw,dassarma}. Now under the fractional magnetic flux, the
eigenvalues for $ n=0, l=0, m=0$ are given by
\begin{eqnarray}
\sqrt{ E_{+,000}^{\uparrow} } = \sqrt{ E_{-,000}^{\downarrow} } = 0,
\end{eqnarray}
and
\begin{eqnarray}
\sqrt{ E_{+,n+100}^{\uparrow} } = \sqrt{ E_{+,n00}^{\downarrow} } =
\sqrt{ E_{-,n+100}^{\downarrow} } = \sqrt{ E_{-,n00}^{\uparrow} }
 = \pm \hbar w_{{l_{B}}} \sqrt{ n +1}
\end{eqnarray}
for $ n \geq 0, l= 0, m=0.$ Here $ w_{l_{B}} \equiv \frac{ \sqrt{2}
v_{{\rm F}} }{l_{B}}$. Equation (70) tells us that there is one
zero-energy state only in the case of up-spin fermions but not in
the case of down-spin fermions at the $ K_{+}$ point. At $
K^{'}_{-}$ point we can have one zero-mode state for down-spin
fermions but not for up-spin fermions. The magnetic field direction
at $ K_{+}$  is opposite to that at $ K^{'}_{-}$. The zero-energy
state may have the four-fold degeneracy emerging from electrons and
holes \cite{pachos}. Since the LL of the zero-energy states becomes
half-filled, no one would observe plateau at $ \nu = 0 $. But by
index theorem, the flux quanta produce $ 4r $ $( r = 0, 1, 2,\cdots
) $ zero-energy states. The $ 2r $ states of these are occupied. The
flux quanta lift the $ 2r $ states to the Fermi energy. And then
they can be removed by doping. The degeneracy between electrons and
holes would be removed. We could observe the Hall plateau at $ \nu =
0 $ because holes are occupied before electrons. And hence we can
describe an experimental observation of the Hall plateau emerging at
$ \nu = 0 $. \cite{kim}

On the basis of the index theorem, we compute the energy spectrum of
the deformed Hamiltonian, $ \mathrm{H}_{+,is} = - 2\hbar^{2} v_{{\rm
F}}^{2}(Q_{+,2is} + Q^{\dag}_{+,2is})^{2} $ at the $ K_{+} $ point.
The up-spin states of zero energy are $ |0>, |1>, |2>, \cdots,
|j^{\uparrow}-1> $. They are degenerate in
$|\psi^{\uparrow}_{+,0ml}>$. On the other hand, for down-spin
states, we have to describe two cases.  As the first case, assume
that $ j^{\downarrow} = 0 $. Then, there do not exist any
zero-energy states. So we may construct the supermultiplet given by
Eq. (68). In the other case, if $ j^{\downarrow} \neq 0, $ the
zero-energy states are given by $ |0>, |1>, |2>, \cdots,
|j^{\downarrow}-1> $ as degenerate states of $
|\psi^{\downarrow}_{+,0ml}>. $ Therefore, these result in the $ (
j^{\uparrow} + j^{\downarrow} )-$fold degeneracy in the zero-energy
state for fermions at the $ K_{+}$ point. This degeneracy implies
the exact correspondence between $ j^{\uparrow} $ fermions and $
j^{\downarrow} $ fermions under deformation. Similarly, we can
investigate the energy spectrum of the zero-energy states at the $
K^{'}_{-} $ point.

In order to generate the up-spin and down-spin states of zero energy
the deformed superoperators are written, in terms of the original
supercharges, by
\begin{eqnarray}
\widehat{Q}_{+,2is} =  \widehat{Q}_{+,2is}^{\dag j^{\downarrow}}
\widehat{Q}_{+,2is}^{ j^{\uparrow}}, & \widehat{Q}_{+,2is}^{\dag} =
\widehat{Q}_{+,2is}^{\dag j^{\uparrow}} \widehat{Q}_{+,2is}^{
j^{\downarrow}},
\end{eqnarray}
where $ j^{\uparrow} $ and $ j^{\downarrow} $ are integers such as $
j^{\uparrow} > j^{\downarrow} $. In terms of the $
\widehat{Q}_{+,2is}^{\dag} $ the state $ | n > $ is given by
\begin{eqnarray}
| n > = \frac{1}{\sqrt{n!}} (\widehat{Q}_{+,2is}^{\dag})^{n} | 0 >.
\end{eqnarray}

Now, let us calculate the deformed eigenvalues by using Eq. (73) and
solving the eigenvalue problems of Eq. (64) for the up-spin and
down-spin components. In the bosonic sector, the deformed energy
eigenvalues are expressed by
\begin{eqnarray}
\sqrt{ E_{+,000}^{\uparrow} } = \sqrt{ E_{-,000}^{\downarrow} } = 0,
\end{eqnarray}
for $ n=0, l=0, m=0. $ And we can have
\begin{eqnarray}
\sqrt{ E_{+,nml}^{\uparrow} }  = \sqrt{ E_{-,nml}^{\downarrow} } =
\pm \hbar w_{{l_{B}}} \sqrt{ \frac{ (n + j^{\uparrow} - 1)!(n +
j^{\downarrow} - 1)! }{ [( n - 1 )!]^{2} } }
\end{eqnarray}
for $ n \geq 1, l = -(j^{\uparrow} - 1), m = 2( j^{\uparrow} - 1 ).
$ In the case of the fermionic sector, if $ j^{\downarrow} = 0 $,
the deformed eigenvalues are
\begin{equation}
\sqrt{ E_{+,nml}^{\downarrow} } = \sqrt{ E_{-,nml}^{\uparrow} } =
\pm \hbar w_{l_{B}} \sqrt{ \frac{ (n + j^{\downarrow} )!(n +
j^{\uparrow})! }{ [( n )!]^{2} } }
\end{equation}
for $ n \geq 0, l = j^{\downarrow}, m=2j^{\downarrow}. $ If $
j^{\downarrow} \neq 0 $, the eigenvalues are given by
\begin{eqnarray}
\sqrt{ E_{+,000}^{\downarrow} } = \sqrt{ E_{-,000}^{\uparrow} } = 0,
\end{eqnarray}
for $ n=0, l=0, m=0. $  And we obtain
\begin{eqnarray}
\sqrt{ E_{+,n+1ml}^{\uparrow} } = \sqrt{ E_{-,n+1ml}^{\uparrow} } =
\pm \hbar w_{l_{B}} \sqrt{ \frac{ (n + j^{\downarrow} - 1 )!(n +
j^{\uparrow} - 1)! }{ [( n - 1 )!]^{2} } }
\end{eqnarray}
for $ n \geq 1, l = j^{\downarrow} - 1, m = 2(j^{\downarrow} - 1). $
We can check up that there exists 4$(j^{\uparrow} +
j^{\downarrow})$-fold degeneracy in the zero-energy states and
4-fold one in all other states.

Among the energy spectrum given above, let us account for the
special cases of $ j^{\uparrow} = 1, $ and $ j ^{\downarrow} = 0 $
and $ j^{\uparrow} = 2, $ and $ j ^{\downarrow} = 0 $. In the case
of $ j^{\uparrow} = 1, $ and $ j ^{\downarrow} = 0 $ it is not hard
to check up that the energy eigenvalues are given by Eqs. (77) and
(78). These results correspond to the spectrum of the monolayer
graphene. For the case of $ j^{\uparrow} = 2, $ and $ j
^{\downarrow} = 0, $ the energy spectra are expressed in terms of
\begin{eqnarray}
\sqrt{ E_{+,000}^{\uparrow} } = \sqrt{ E_{-,000}^{\downarrow} } = 0,
\end{eqnarray}
for $ n=0, l = 0, m=0. $ And we can obtain \cite{ezawa}
\begin{equation}
\sqrt{ E_{+,nml}^{\uparrow} } = \sqrt{ E_{-,nml}^{\downarrow} } =
\pm \hbar w_{l_{B}} \sqrt{ n(n +1) }
\end{equation}
for $ n \geq 1, l = -1, m = 2, $ while having
\begin{equation}
\sqrt{ E_{+,nml}^{\downarrow} } = \sqrt{ E_{-,nml}^{\uparrow} } = =
\pm \hbar w_{l_{B}} \sqrt{ (n +1)(n + 2) }
\end{equation}
for $ n \geq 0, l = 0, m = 0. $ These energy spectra are eigenvalues
of the bilayer graphene affected by the deformation, and are in
agreement with the results in the literature
\cite{kim,mccann,feldman}. On the basis of the index theorem we have
shown that there exist the $ j^{\uparrow}$-fold and $
j^{\downarrow}$-degeneracy in the zero-energy state at the $ K_{+} $
point and similarly at the $ K^{'}_{-} $. And hence we can obtain
the QHE characterized by \cite{park}
\begin{eqnarray}
\sigma_{xy} = \nu \frac{ e^{2} }{ h }, &  \nu =\pm 4( |n| +
\frac{j^{\uparrow} + j^{\downarrow}}{2} ).
\end{eqnarray}

\section{\label{sec:level1} Summary and conclusion }

We presented the electronic properties of massless Dirac fermions
characterized by geometry and topology on a graphene sheet in this
chapter. Topological effects can be elegantly described by the
Atiyah-Singer index theorem. It provides a topological invariant
under deformations on the Dirac operator and plays an essential role
in formulating supersymmetric quantum mechanics over twisted
Dolbeault complex associated with the deformation on the topology of
the lattice in a graphene system. Exploiting the G-index theorem and
a high degree of symmetry, we explained deformed energy eigenvalues
in graphene. The Dirac fermions result in SU(4) symmetry emerging
out of both the pseudospin and spin as a high degree of symmetry in
the noninteracting Hamiltonian of monolayer graphene. Under the
topological deformation the zero-energy states emerge naturally
without the Zeeman splitting at the Fermi points in the graphene
sheet. Thus we observed an emergence of a higher degree of hidden
symmetry under the topological deformation in graphene while the
pseudospin is a good symmetry at the $ K $ and $ K^{'} $ points in
graphene. In the particular SU(2) of the pseudospin, the SU(2) is
the exact spin symmetry of each Landau level. In the case of nonzero
energy, the up-spin and down-spin states have the exact high
symmetries of spin, forming the pseudospin singlet pairing. The
pseudospin can play a key role on the physics of the $ n = 0 $ LL in
the graphene sheet. The valley pseudospin degeneracy can lift only
at the zeroth LL. The 4-fold degeneracy can be removed in the
zero-energy states of monolayer graphene. If the mass terms were
taken into account, the four-fold degeneracy can be removed in the
zero-energy state of monolayer graphene. We can exploit this to
understand the emergence of a Hall plateau at $ n = 0 $ in the
experimental observations. But the four-fold degeneracy is not
removed in the higher LLs. Including the Coulomb interaction, we can
lift the degeneracy. The pseudospin symmetry $\mathrm{SU}(2)$ is
broken to $\mathrm{U}(1)\times \mathrm{Z}_{2}.$ Therefore the total
symmetry gives rise to $ \mathrm{SU}(2)_{spin} \times( \mathrm{U}(1)
\times \mathrm{Z} )_{psuedospin} $ while the spin symmetry $
\mathrm{SU}(2)$ remains to be exact. Hence we understood the
peculiar and unconventional quantum Hall effects of the $ n = 0 $
Landau level in monolayer graphene on the basis of the index theorem
and the high degree of symmetry under the topological deformation
without the Zeeman splitting. It would be very interesting and quite
possible to apply the present approach to investigation of the
composite Dirac fermions and fractional quantum Hall effects in
graphene.

\begin{acknowledgments}
\noindent
This work was supported by the Korea Science and
Engineering Foundation through the National Research Laboratory
Program (R0A-2005-001-10152-0), by Priority Research Centers Program
through the National Research Foundation of Korea (NRF) funded by
the Ministry of Education, Science and Technology (2009-0094037),
and by the Brain Korea 21 Project.
\end{acknowledgments}

\bibliography{amo,mag}

\end{document}